\documentclass[doublecol]{epl2} 
\usepackage[switch]{lineno}

\title{Experimental investigation into Lagrangian statistics of droplets in homogeneous isotropic turbulence}
\shorttitle{Drops in HIT} 

\author{Lu Li\inst{1} \and Yi-Bao Zhang\inst{1} \and Yaning Fan\inst{1} \and  Federico Toschi \inst{2}  \and Chao Sun\inst{1,3} \thanks{E-mail:\email{chaosun@tsinghua.edu.cn}}}
\shortauthor{L. Li \etal}

\institute{                    
  \inst{1} New Cornerstone Science Laboratory, Center for Combustion Energy, Key Laboratory for Thermal Science and Power Engineering of MoE, Department of Energy and Power Engineering, Tsinghua University, 100084 Beijing, China\\
  \inst{2} Department of Applied Physics and Science Education, Eindhoven University of Technology, 5600 MB Eindhoven, The Netherlands\\
  \inst{3} Department of Engineering Mechanics, School of Aerospace Engineering, Tsinghua University, 100084 Beijing, China
}
\pacs{nn.mm.xx}{First pacs description}
\pacs{nn.mm.xx}{Second pacs description}
\pacs{nn.mm.xx}{Third pacs description}

\abstract{We experimentally investigate the Lagrangian dynamics of finite-sized, neutrally buoyant droplets in homogeneous isotropic turbulence. The droplet size follows a log-normal distribution whose average value decreases with increasing Reynolds number, reflecting enhanced turbulent breakup. While size-conditioned velocity and acceleration statistics show only weak finite-size dependence, temporal measures reveal clear size-dependent dynamics: larger droplets exhibit longer Lagrangian velocity integral times and an extended ballistic regime in their mean squared displacement. These findings indicate that though droplets exhibit mild deformation and internal circulation, they behave similarly to finite-size rigid particles in terms of Lagrangian dynamics. Our study opens the way to study droplet-laden turbulence and droplet-flow interactions.}

\begin{document}

\maketitle

\section{Introduction}

Droplet-laden flows play a central role in numerous natural and industrial processes, including diesel combustion \cite{post2002modeling,sirignano2010fluid}, mixing and evaporation in chemical engineering \cite{martinez2023vapor,ren2017thermal}, rain formation in warm clouds \cite{pinsky1997turbulence,falkovich2002acceleration,shaw2003particle}, air-quality and pollution dynamics \cite{seinfeld1986atmospheric,csanady2012turbulent}and plankton transport in the ocean \cite{rothschild1988small,lewis2000planktonic}. In such systems, the carrier phase is typically turbulent, exhibiting a hierarchy of eddies that that span from large, energy-containing scales down to the Kolmogorov scale where dissipation dominates \cite{frisch1995turbulence,davidson2015turbulence}. Subjected to turbulent strain and shear, suspended droplets continuously deform, coalesce, and break up, generating a cascade of fragments across a wide range of sizes \cite{elghobashi2019direct,mathai2020bubbly,ni2024deformation,rodriguez2025drop}. This intrinsic polydispersity adds substantial complexity to the dynamics and presents longstanding challenges for theoretical modeling, multiscale simulation, and experimental characterization.

When droplets are sufficiently small and neutrally buoyant, they behave as passive tracers that closely follow the local fluid motion \cite{voth2002measurement,toschi2009lagrangian}. As their size increases, however, finite inertia causes them to deviate from the surrounding flow \cite{bec2006acceleration,volk2008acceleration,fan2024accelerations}. For small, rigid particles that are much heavier than the carrier fluid, this inertial effect is commonly characterized by the Stokes number, $St=\tau_d/\tau_\eta$, which compares particle response time with the Kolmogorov timescale. For finite $St$, particles exhibit reduced acceleration variance and longer Lagrangian velocity correlation time \cite{voth2002measurement,qureshi2007turbulent,calzavarini2009acceleration}. Their spatial distribution also becomes density-dependent: light bubbles accumulate in vortical regions \cite{deng2006bubble,reuter2017flow}, while heavy particles cluster in high-strain zones \cite{squires1991preferential,eaton1994preferential,bec2005clustering}. Such clustering strongly influences interactions among dispersed phases and plays a key role in processes such as rainfall initiation \cite{vaillancourt2000review,falkovich2002acceleration,shaw2003particle}.

Finite-size effects arise not only from inertia but also from the fact that larger droplets sample the curvature of flow across their own spatial extent. To capture these effects, Fax\'en and Gatignol introduced corrections to extend the classical Maxey-Riley equation by incorporating higher-order flow gradients \cite{faxen1922widerstand,gatignol1983faxen,annamalai2017faxen,dolata2021faxen}. These terms account for the spatial averaging experienced by a finite-sized particle and predict a reduction in acceleration variance relative to point tracers \cite{voth2002measurement,qureshi2007turbulent,calzavarini2009acceleration}. Notably, the same equation--without the Fax\'en corrections--was independently derived by Shu-Tang Tsai in 1957 and published in Chinese (see \cite{tsai2022sedimentation} for recent translation). Over the past two decades, numerous experimental and numerical studies have examined the acceleration statistics of finite-size particles, validating such theoretical predictions and elucidating the roles of inertia and turbulence intermittency \cite{balkovsky2001intermittent,falkovich2004intermittent,sundaram1997collision,collins2004reynolds,xu2008motion,naso2010interaction}. However, these efforts have primarily focused on rigid spheres or nearly neutrally buoyant bubbles with tightly controlled size distributions \cite{qureshi2007turbulent,mathai2018dispersion}, conditions that differ markedly from the inherently polydisperse droplet populations found in emulsions and industrial flows.

In contrast to these controlled-size, single-phase studies, our experiment allows the turbulence itself to generate a naturally polydisperse droplet population. This enables simultaneous measurement of the droplet-size distribution and size-conditioned Lagrangian dynamics. Achieving this requires long-duration, three-dimensional tracking of individual droplets that resolves both small-scale fluctuations and multi-timescale statistics--an experimental challenge in turbulent multiphase flows. Our rotating-propeller facility produces highly homogeneous and isotropic turbulence, providing a robust platform for probing finite-size effects. Combined with a state-of-the-art PTV system, it enables high-fidelity Lagrangian tracking that clearly reveals the influence of finite size on droplet dynamics. Owing to its similarity to swirling mixing chambers widely used in industrial applications, our setup also offers insights relevant to processes involving mixing, dispersion, and droplet-flow interactions.

\section{Experimental setup}

\begin{figure*}[ht]
\onefigure[width=0.8\textwidth]{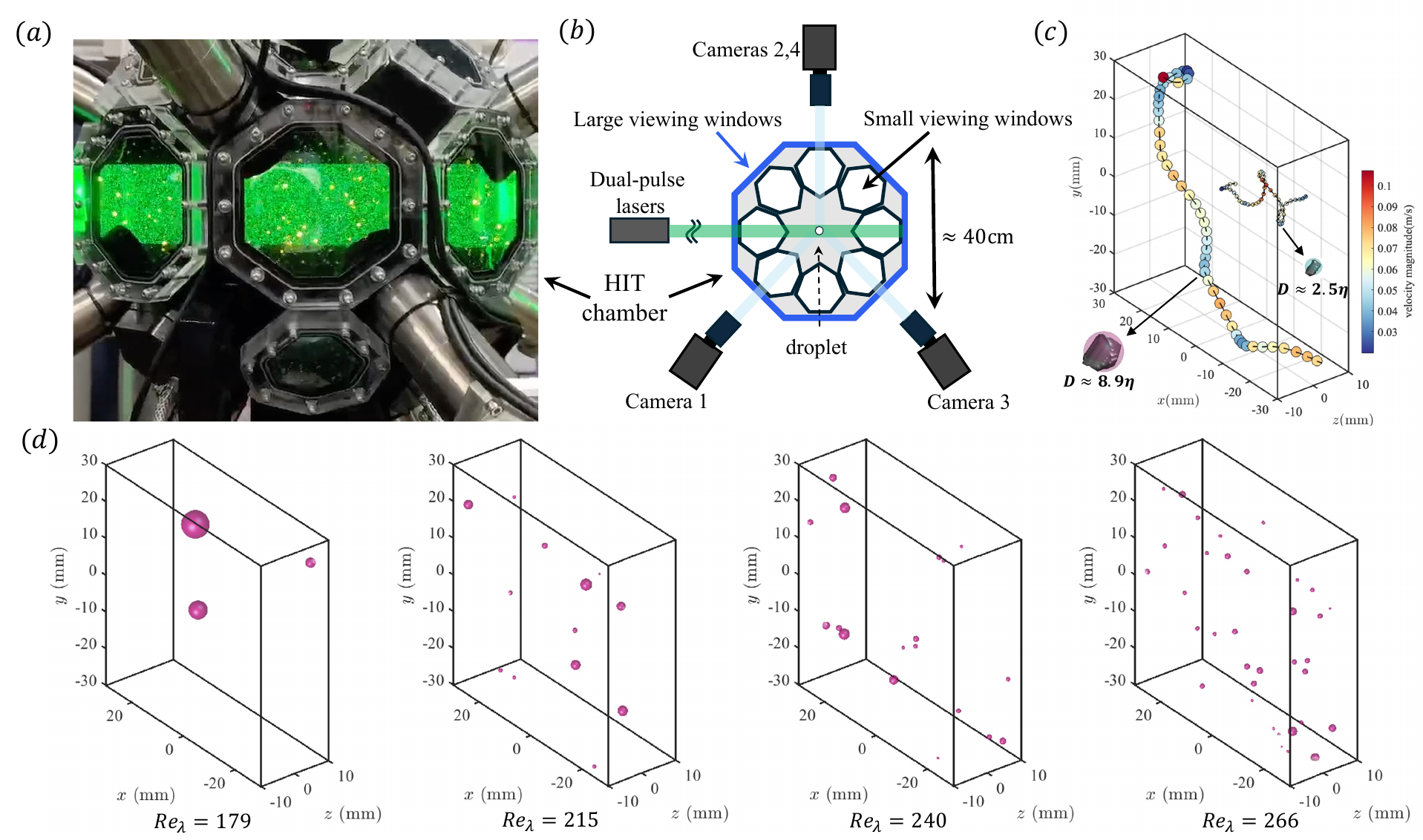}
\caption{Experimental setup for droplet Lagrangian tracking. 
(a) Photograph of the experiment conducted in the ``soccer-ball-like" HIT chamber. 
(b) Schematic of the optical arrangement from a plan view: four high-speed cameras capture droplet motion from different viewing angles to enable accurate three-dimensional reconstruction. Three cameras are mounted on the same horizontal plane and view the flow through the large windows, while a fourth camera is tilted at $45^\circ$ to view through the smaller windows. All cameras are synchronized with a double-cavity high-speed laser that generates a $\approx 2$ cm thick pulsed illumination volume. 
(c) Reconstructed three-dimensional trajectories at $Re_\lambda \approx 215$, color-coded by the instantaneous velocity magnitude. The trajectories span 550 consecutive frames, corresponding to a physical duration of 1.83 s ($\approx 87\tau_\eta$), where $\tau_\eta$ is the Kolmogorov timescale. Insets illustrate the determination of the droplet effective diameter from the visual-hull-based voxel reconstruction for a small droplet ($D \approx 2.5\eta$) and a large droplet ($D \approx 8.9\eta$). 
(d) Variation of droplet size and number density with increasing Taylor-Reynolds number $Re_\lambda$. For clarity, droplet sizes are scaled by a factor of two.}
\label{fig.1}
\end{figure*}

The experiments were performed in the central region of a closed, ``soccer-ball-like" facility, similar in design to the Lagrangian Exploration Module used in previous studies \cite{zimmermann2010lagrangian,wang2025lagrangian}. A photograph of the setup is shown in Fig. \ref{fig.1}(a). The supporting framework is made of stainless steel, while the side panels, comprising both large and small sections, are constructed from high-transparency acrylic glass to ensure unobstructed optical access. The effective diameter of the flow chamber is approximately 40 cm, corresponding to a total volume of about 30 L. Turbulence is generated by twelve propellers mounted on the chamber vertexes, each independently driven by an AC servo motor (MS1H4-40B30CB, INOVANCE) via a planetary gearbox (VRB-060C-3-K3-14BK14, Nidec). The bulk turbulence properties were characterized using two-dimensional PIV measurements, and the key parameters are summarized in Table \ref{tab.1}.

\begin{table}[ht]
\caption{A summary of the bulk flow characteristic. $f$ denotes the propeller rotation rate, $Re_\lambda$ is the Taylor-Reynolds number, $u^\prime$ is the root-mean-squared fluctuation velocity, $\epsilon$ is the energy dissipation rate, measured from the inertial range scaling of the Eulerian velocity structure functions, and $\eta$ and $\tau_\eta$ are the Kolmogorov length and timescales, respectively $\lambda$ stands for Taylor microscale.}
\label{tab.1}
\begin{center}
	\begin{tabular}{ccccccc}
      	\hline\hline
		$f$    & $Re_\lambda$ & $u^\prime$  & $\epsilon$  & $\tau_\eta$ & $\eta$      & $\lambda$\\
		rpm&              & m/s& ${\rm m}^2/{\rm s}^3$& ms& $\mu$m&  mm\\
		\hline
		100	& 179 & 0.033   & $6.18\times 10^{-4}$ 	&38     &184 	&4.83\\
		150	& 215 & 0.049	& $2.07\times 10^{-3}$ 	&21     &136	&3.91\\
		200	& 240 & 0.065	& $5.12\times 10^{-3}$ 	&13     &108	&3.29\\
		250	& 266 & 0.079	& $9.32\times 10^{-3}$ 	&10     &93		&3.00\\
		300	& 287 & 0.092	& $1.48\times 10^{-2}$ 	&8      &83		&2.78\\
		\hline\hline
   \end{tabular}
\end{center}
\end{table}

During the droplet-tracking experiments, all propellers were driven at the same constant frequency so that the flow in the central measurement region is  approximately homogeneous and isotropic \cite{zimmermann2010lagrangian}. To generate droplets, approximately 30 mL of silicone oil (Shin-Etsu KF-99; density $\rho_d \approx 1\times 10^3$ kg/m$^3$, kinematic viscosity $\nu_d \approx 2.0\times 10^{-5}$ m$^2$/s, surface tension $\gamma_d \approx 20$ mN/m), doped with Nile Red dye for fluorescence imaging, was injected into the water-filled chamber (density $\rho_c \approx 1\times 10^3$ kg/m$^3$, kinematic viscosity $\nu_c \approx 0.89\times 10^{-6}$ m$^2$/s, surface tension $\gamma_c \approx 72$ mN/m) at a rate of 120 mL/min through a 2-mm inner-diameter stainless-steel needle using a syringe pump (PHD Ultra, Harvard Apparatus), yielding a dispersed droplet population with an initial characteristic diameter of approximately 4 mm and a resulting volume fraction of about 0.1\%. Such a low volume fraction of the dispersed phase places the system in a one-way coupling regime, similar to pervious studies\cite{elghobashi1994predicting,mathai2018dispersion}, where the background turbulence remains largely unaffected. A 2-cm-thick laser volume was formed to illuminate the droplets in the central measurement region, enabling high-contrast imaging of their positions and shapes. A representative photograph of the illuminated droplets, acquired at a reduced propeller rotation rate of $f = 100$ rpm (corresponding to $Re_\lambda = 179$), is shown in Fig. \ref{fig.1}(a). The complete optical arrangement is sketched in Fig. \ref{fig.1}(b). Four high-speed cameras (Photron NOVA S12, each equipped with a 100-mm lens) were arranged around the chamber to record the droplet motion from different viewing angles. Each camera was also equipped with a narrow-band optical filter (570 nm center wavelength, 20 nm bandwidth, OD5) to suppress scattered light from out-of-plane droplets, thereby reducing image overlap and improving droplet detection. Depending on the propeller rotation rate, the cameras with a resolution of 1024 $\times$ 1024 pixels recorded at high frame rates, providing synchronized multi-view measurements for three-dimensional Lagrangian tracking.

Droplet trajectories were reconstructed from multi-camera recordings using a ray-traversal PTV algorithm \cite{mathai2018dispersion,bourgoin2020using}, which imposes no lower particle-size limit and thus enables robust tracking of finite-size droplets. Figure~\ref{fig.1}(c) presents representative trajectories, and the insets illustrate the determination of the sphere-equivalent diameter $D = (6V/\pi)^{1/3}$ from the reconstructed volume $V$ obtained using four silhouette images (gray objects in Fig.~\ref{fig.1}(b)). Small droplets ($D \approx 2.5\eta$) exhibit strongly curved, swirling trajectories, whereas larger droplets ($D \approx 8.9\eta$) follow smoother paths, reflecting the increased influence of inertia on their dynamics. The observed droplet poly-dispersity arises naturally in the HIT chamber due to the action of the rotating propellers. As shown in Fig.~\ref{fig.1}(d), higher $Re_\lambda$ conditions generate smaller droplets with higher number densities. Notably, in each experiment, the flow was allowed to evolve for 5 min after oil injection to ensure statistical stationarity before imaging started. The memory of each camera limited recordings to 21,000 frames, corresponding to several large-eddy turnover times, $t_E = {u^\prime}^2/\epsilon$, sufficient to obtain converged statistics. In addition to the droplet-tracking experiments, reference tracer measurements were conducted at the same sampling frequency and propeller rotation rate.

\section{Droplet size distribution}
\begin{figure}
\onefigure[width=0.85\columnwidth]{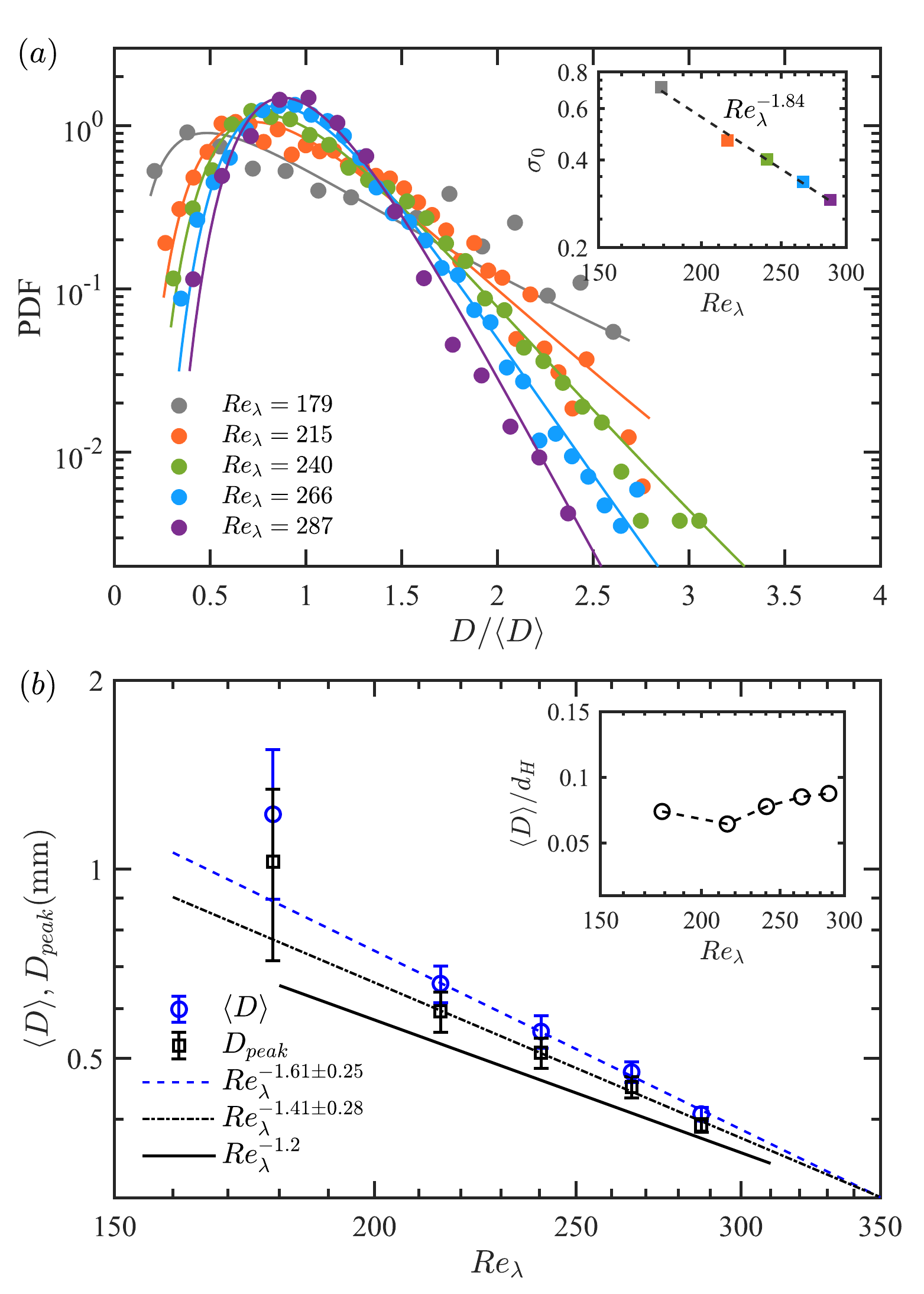}
\caption{
(a) PDFs of the normalized droplet diameter, $D / \langle D \rangle$, for a range of $Re_\lambda$. Solid lines denote the corresponding log-normal fits. Each PDF is constructed from approximately $\mathcal{O}(10^3)$ droplet trajectories. The inset shows the fitted standard deviation $\sigma_0$ plotted against $Re_\lambda$. 
(b) Mean droplet diameter $\langle D \rangle$ and the most probable diameter $D_{peak}$ as functions of $Re_\lambda$. The corresponding best-fit trends are shown by the blue dashed line and the black dash-dotted line, respectively.  For comparison, the Kolmogorov-Hinze scaling $\langle D \rangle \sim Re_\lambda^{-1.2}$ is also included. The inset displays the ratio of $\langle D \rangle$ to the Hinze scale $d_H$ (Eq. \ref{eq:Hinze}).
}
\label{fig.2}
\end{figure}
As a key feature of turbulent emulsions, the size distribution of dispersed droplets governs both their microscopic rheological behavior and the macroscopic transport properties of the flow \cite{mason1999new,wang2019self,yi2021global,wang2022finite,wang2022turbulence,su2025interfacial}. Under the present stationary stirring conditions and at low volume fractions, the droplets evolve toward a statistically steady size distribution. As the Reynolds number increases, this distribution shifts toward smaller sizes, since stronger shear enhances droplet deformation and promotes breakup. This trend is already evident qualitatively in Fig.~\ref{fig.1}(d), where the droplet sizes decrease with increasing $Re_\lambda$. A more quantitative assessment is obtained from the effective droplet diameter computed via voxel-space reconstruction. It should be noted that, although shear can deform droplets, such deformations are not considered here; droplets are characterized solely by their effective diameters.

Figure~\ref{fig.2}(a) shows the probability density functions (PDFs) of droplet size for different $Re_\lambda$, where the normalized diameter is defined as $\tilde{D} = D/\langle D\rangle$, and $\langle D\rangle$ denotes the ensemble-averaged diameter. For the higher $Re_\lambda$ cases, the PDFs are well described by a log-normal distribution,
\begin{equation}
	P(\tilde{D}) = \frac{1}{\tilde{D}\sigma_0\sqrt{2\pi}}
	\exp\left[-\frac{\left(\log\tilde{D}-\log\tilde{D}_0\right)^2}{2\sigma_0^2}\right]
\end{equation}
where $\tilde{D}_0$ and $\sigma_0$ are fitting parameters. In contrast, the lowest $Re_\lambda = 179$ case exhibits slight deviations from the log-normal form, showing a double-peaked structure reminiscent of the droplet-size distributions reported in two-phase Taylor-Couette flows \cite{fan2025experimental}.

Despite this exception at low $Re_\lambda$, the predominance of log-normal distributions at higher $Re_\lambda$ suggests that droplet fragmentation is the dominant mechanism governing droplet formation in the present system. This observation is consistent with previous studies on atomization processes \cite{villermaux2020fragmentation}, raindrop breakup \cite{feingold1986lognormal}, and droplet-laden turbulence \cite{yi2021global,yi2023recent}. Moreover, the fitted standard deviation $\sigma_0$ decreases monotonically with increasing $Re_\lambda$, $\sigma_0\sim Re_\lambda^{-1.84}$ (inset of Fig.~\ref{fig.2}(a)), indicating that stronger turbulence leads to narrower size distributions and more monodisperse droplets \cite{yi2021global}.

To further investigate how droplet size depends on turbulence intensity, Fig. \ref{fig.2}(b) shows both the mean droplet diameter, $\langle D\rangle$, and the most probable diameter, $D_{peak}$, as functions of the Taylor-Reynolds number, $Re_\lambda$. The mean diameter is calculated directly from the raw data, whereas the most probable diameter is derived from log-normal fits. Power-law fits (applied only for $Re_\lambda > 179$ due to the the limited number of recorded droplets at $Re_\lambda = 179$) yield scaling exponents of $-1.61 \pm 0.25$ for $\langle D\rangle$ and $-1.41 \pm 0.28$ for $D_{peak}$. Both of them are in reasonable agreement with the theoretical prediction of $-1.20$ (within the experimental uncertainty), obtained by substituting the energy dissipation rate $\epsilon$ into the Kolmogorov-Hinze relation \cite{Kolmogorov1949,hinze1955fundamentals,perlekar2012droplet}:
\begin{equation}\label{eq:Hinze}
d_H = \left(\frac{We_c}{2}\right)^{5/3}\left(\frac{\rho_c}{\gamma}\right)^{-3/5}\epsilon^{-2/5} \sim Re_\lambda^{-6/5}
\end{equation}
where $We_c$ is the critical Weber number, and the interfacial tension between the two phases can be estimated via $\gamma = \gamma_c + \gamma_d - 2\xi(\gamma_c \gamma_d)^{1/2}$ \cite{lee1993scope,fan2025experimental}. In our case, with the interfacial interaction parameter $\xi = 1$, $\gamma \approx 16$ mN/m. 

It should be emphasized, however, that the observed scaling does not imply that droplet breakup predominantly occurs in the bulk. Indeed, the measured mean diameter $\langle D\rangle$ is much smaller than the Hinze scale $d_H$ estimated from the bulk energy dissipation rate $\epsilon$; as shown in the inset of Fig.~\ref{fig.2}(b), the ratio $\langle D\rangle / d_H$ is less than 0.1. Instead, breakup primarily occurs near the rotating propellers, where the local strain rate far exceeds that in the bulk \cite{stelmach2023mixing,ni2024deformation}, as commonly observed in wall-bounded turbulence \cite{grossmann2016high,su2024numerical,zhang2025global}. Although the dynamics in the propeller region are of interest, they cannot be measured due to the restriction of the optical windows of the setup.

Finally, to the best of our knowledge, no existing theoretical framework quantitatively predicts the scaling of $\sigma_0$ with $Re_\lambda$. Nevertheless, under the framework of sequential cascades of breakups \cite{yi2021global}, both the mean $\langle D\rangle$ and the variance $\langle D^2\rangle$ of the droplet-size distribution decrease with increasing $Re_\lambda$. Consequently, $\sigma_0$, which reflects the relative width of the distribution, is also expected to follow a decaying power law with increasing Reynolds number.

\section{Lagrangian statistics}
After establishing the droplet size distribution, we next examine how droplet size influences their Lagrangian statistics and dynamics by analyzing various size-conditioned quantities. The results are qualitatively similar across different $Re_\lambda$, with those for $Re_\lambda=266$ shown below.

\begin{figure}
\onefigure[width=0.85\columnwidth]{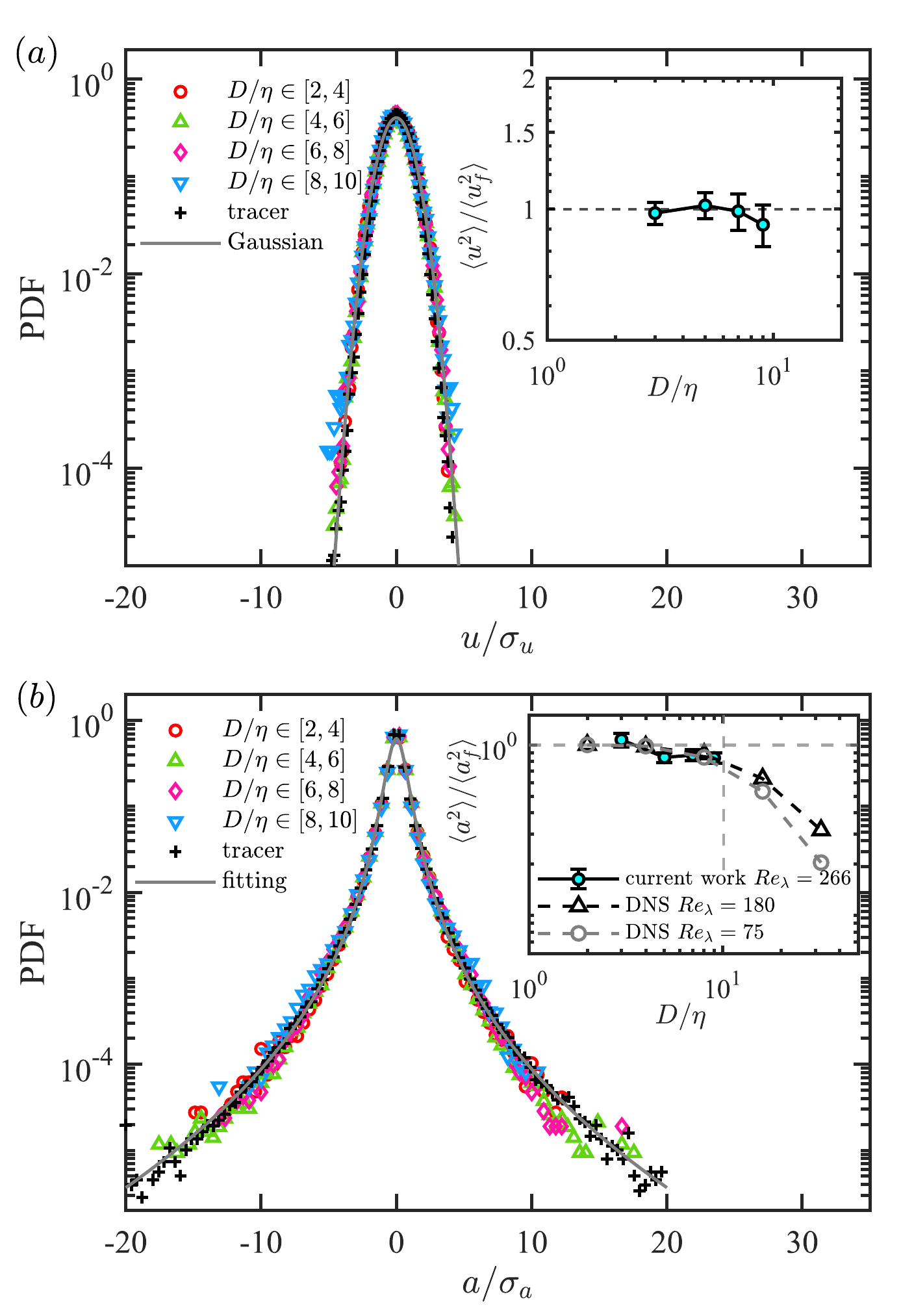}
\caption{PDFs of (a) normalized velocity $u/\sigma_u$ and (b) normalized acceleration $a/\sigma_a$ for droplets at $Re_\lambda = 266$, conditioned on size ranges $D/\eta \in [2,4]$, $D \in [4,6]$, $D \in [6,8]$, and $D \in [8,10]$, with the results of tracer experiment at the same $Re_\lambda$ as well as the Gaussian distribution and stretched exponential fit as references. In both panels, the inset shows the ratio of velocity variance and acceleration variance to those of tracer. In the inset of panel (b), Fax\'en-corrected numerical results at $Re_\lambda = 75$ and $Re_\lambda = 180$ \cite{calzavarini2009acceleration} are included.}
\label{fig.3}
\end{figure}

Figure \ref{fig.3}(a) shows the PDFs of normalized velocity and acceleration for Lagrangian trajectories conditioned on different droplet-size ranges. The PDFs of conditioned velocity nearly coincide with those of tracers at the same $Re_\lambda$, exhibiting an approximately Gaussian shape. The inset of Fig.~\ref{fig.3}(a) compares the velocity variance of finite-size droplets ($\langle u^2\rangle$) with that of tracers ($\langle u_f^2\rangle$). The ratio $\langle u^2\rangle / \langle u_f^2\rangle$ remains close to unity across all droplet sizes, with only a slight reduction for larger droplets. This trend aligns with the asymptotic prediction $\langle u^2\rangle - \langle u_f^2\rangle = -(\langle u_f^2\rangle/100)(D/\lambda)^2$ \cite{homann2010finite}. Although our limited range of $D/\lambda < 1/3$ excludes direct quantitative verification, the expected monotonic decrease is evident.

Figure \ref{fig.3}(b) shows that the PDFs of normalized acceleration for both finite-size droplets and tracers nearly collapse onto a single curve, which is well described by the stretched-exponential form \cite{mordant2004experimental,wang2025lagrangian}:
\begin{equation}\label{eq:Acc_PDF}
P(\tilde{a}) = C \exp\left(
    -\frac{\tilde{a}^2}{
        \left(1 + \left(\frac{|\tilde{a}|\beta}{\sigma}\right)^\gamma\right)\sigma^2}
    \right),
\end{equation}
where $\tilde{a} = a/\sigma_a$, $\beta = 0.49$, $\gamma = 1.59$, $\sigma = 0.65$, and $C = 0.62$ is a normalization constant. A closer inspection of the tail region, however, reveals slight deviations between droplets and tracers: the PDF tails for droplets are less pronounced, indicating that finite-size droplets act as spatial filters that smooth out the smallest turbulent fluctuations and reduce Lagrangian intermittency. The inset of Fig.~\ref{fig.3}(b) shows the droplet-size-conditioned acceleration variance, normalized by the tracer acceleration variance $\langle a_f^2\rangle$, as a function of normalized diameter $D/\eta$. For comparison, numerical simulations at $Re_\lambda = 75$ and $180$ \cite{calzavarini2009acceleration} are included. Under this normalization, all datasets collapse onto an almost universal curve, suggesting that Fax\'en-type corrections \cite{faxen1922widerstand,annamalai2017faxen,dolata2021faxen} also capture the size dependence of liquid droplets by accounting for finite-size filtering of small-scale velocity gradient \cite{bec2006acceleration,volk2008acceleration,calzavarini2009acceleration}. This agreement implies that liquid droplets, despite their interfacial tension and internal viscosity, experience an effective forcing similar to that of rigid, neutrally buoyant solid particles, at least for moderate $D/\eta$.

It is important to note that the present measurements are confined to droplets with $D/\eta < 10$, which limits the scale separation between the droplet diameter and the Kolmogorov length scale. Within this regime, the droplet Reynolds number remains relatively small, and the dominant finite-size influence manifests primarily as a suppression of the most intense dissipative-scale fluctuations. Consequently, the resulting Lagrangian statistics deviate only mildly from those of tracer particles. This observation, however, should not be interpreted as evidence that finite-size effects exert only a minor influence on the underlying Lagrangian dynamics. As demonstrated in the following section, these dynamical effects can be considerably more pronounced.
 
\section{Lagrangian dynamics}
\begin{figure}
\onefigure[width=0.85\columnwidth]{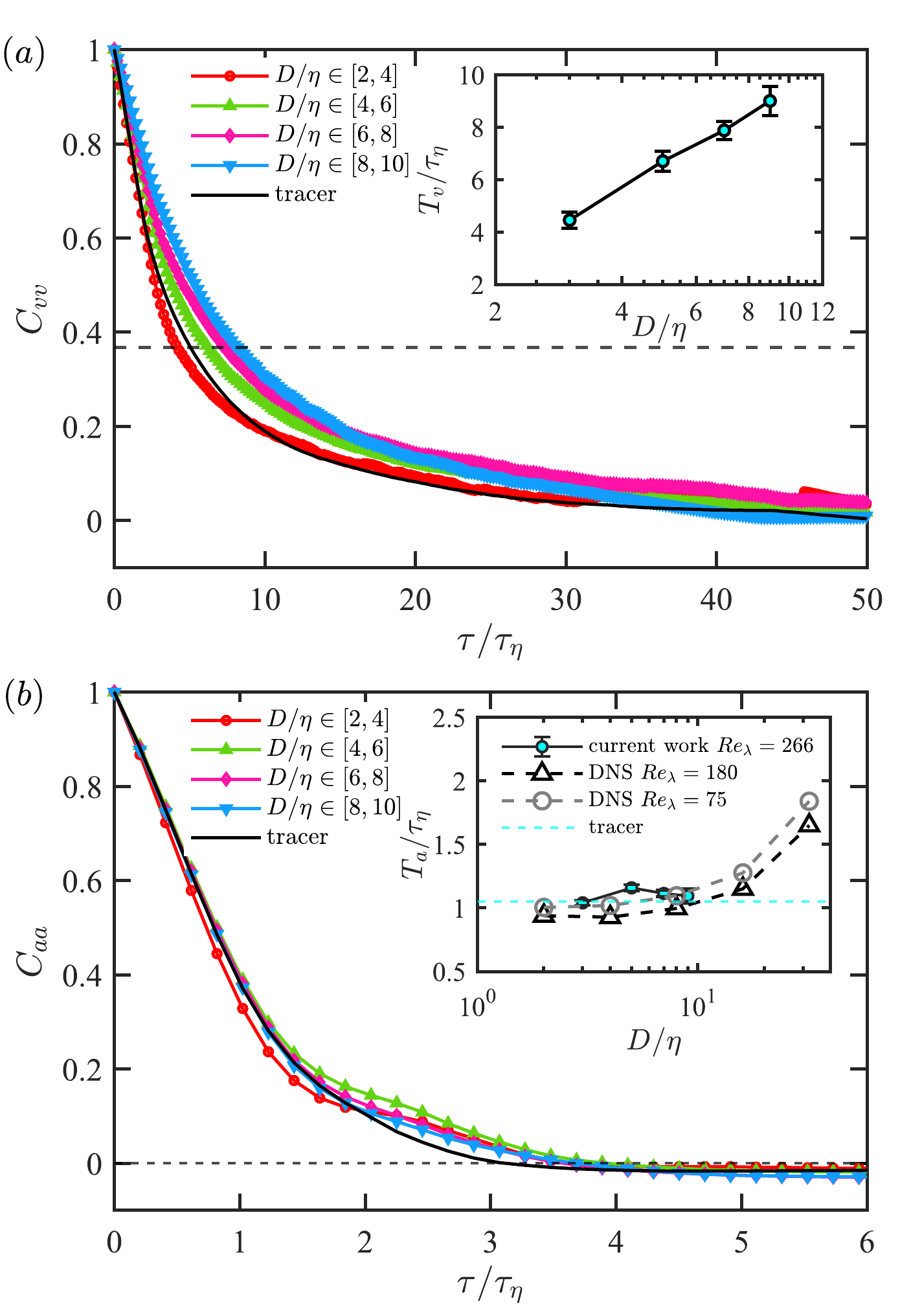}
\caption{(a) Velocity ACF $C_{\boldsymbol{vv}}(\tau)$and (b) acceleration ACF $C_{\boldsymbol{aa}}(\tau)$ at $Re_\lambda = 266$, conditioned on four droplet-size ranges: $D \in [2,4]$, $D \in [4,6]$, $D \in [6,8]$, and $D \in [8,10]$. Tracer results at the same $Re_\lambda$ are shown for reference. Insets show Lagrangian integral times: $T_v$ is defined as the characteristic time at which $C_{\boldsymbol{vv}}(\tau)$ decays to $1/e$, and
$T_a = \int_0^{T_0} C_{\boldsymbol{aa}}(\tau)\,\mathrm{d}\tau$ with $T_0$ the first zero-crossing of $C_{\boldsymbol{aa}}$.
Fax\'en-corrected numerical results at $Re_\lambda = 75$ and $Re_\lambda = 180$\cite{calzavarini2009acceleration} are included for comparison.}
\label{fig.4}
\end{figure}
To assess how finite-size effects shape the temporal Lagrangian dynamics of droplets, we examine the velocity and acceleration autocorrelation functions (ACFs) and the mean squared displacement (MSD), which quantify the size-dependent response to turbulent forcing. The ACFs and MSD are defined as:
\begin{equation}
    C_{\bm{qq}}(\tau)=\frac{\langle \bm{q}(t)\cdot\bm{q}(t+\tau)\rangle}{\langle |\bm{q}(t)|^2\rangle}
\end{equation}
\begin{equation}
    \langle |\Delta X|^2\rangle(\tau)=\langle |\bm{x}(t+\tau)-\bm{x}(t)|^2\rangle
\end{equation}
where $\bm{q}$ denotes either the velocity $\bm{u}$ or acceleration $\bm{a}$, and $\bm{x}$ is the droplet position.

Figure \ref{fig.4}(a) and (b) show the velocity and acceleration ACFs for different droplet-size ranges. The influence of droplet size is far more pronounced in the velocity ACFs. As the droplet diameter increases, the velocity ACF progressively departs from that of the tracer and shifts toward longer times, leading to an increased normalized integral time $T_v/\tau_\eta$, as seen in the inset. The slight discrepancy between the tracer data and the smallest droplet-size class can be attributed to statistical uncertainty and to differences in the tracking methods: tracer trajectories were obtained using a shake-the-box algorithm, whereas droplet trajectories were processed using an in-house ray-traversal PTV algorithm. Despite this, the size dependence of the velocity ACF is unambiguous.

In contrast, the acceleration ACFs exhibit minimal variation across droplet sizes and nearly collapse onto a single curve. No systematic trend is observed in the normalized acceleration integral time $T_a/\tau_\eta$. This behavior is consistent with the Fax\'en-corrected numerical predictions shown in the inset of Fig.~\ref{fig.4}(b), which also display negligible growth in $T_a/\tau_\eta$ for droplets with $D/\eta < 10$. Taken together, these results indicate that velocity fluctuations, quantified through the acceleration, decorrelate rapidly, while the velocity itself retains memory for increasingly long times as droplet size grows. In other words, larger droplets tend to maintain their velocity over longer durations than smaller ones, suggesting an extended ballistic regime for larger droplets.

\begin{figure}
\onefigure[width=0.95\columnwidth]{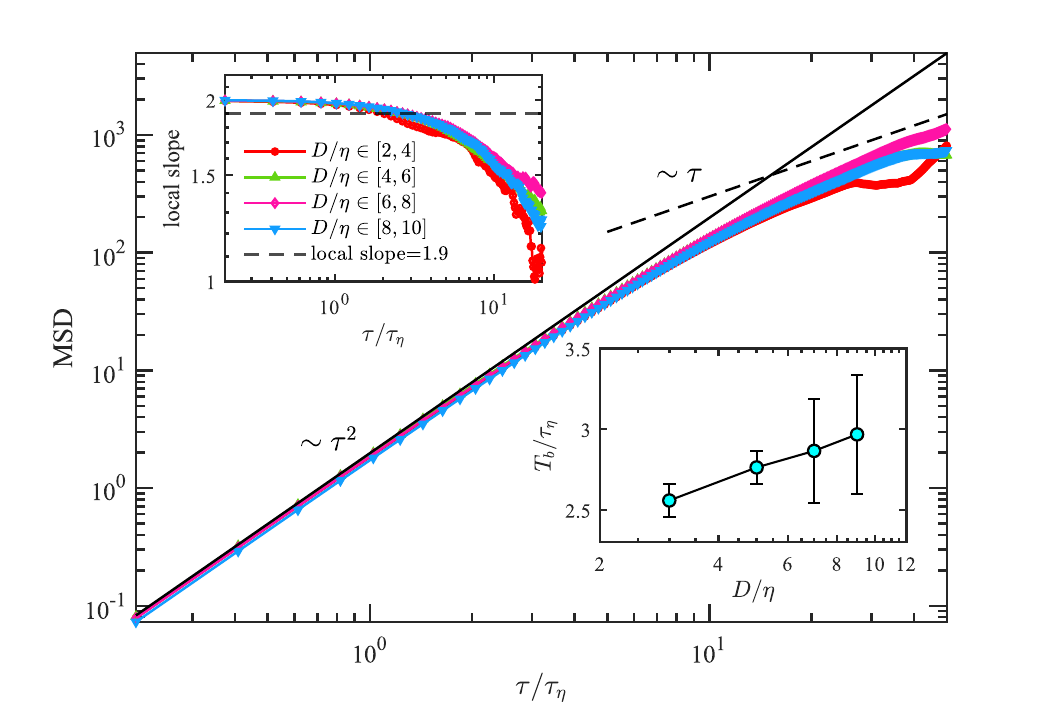}
\caption{MSD of droplets with different diameters at $Re_\lambda=266$. The MSD exhibits an initial ballistic regime, $\langle |\Delta X|^2 \rangle \sim \tau^2$ (black solid line), and then progressively transitions toward a diffusive regime, $\langle |\Delta X|^2 \rangle \sim \tau$ (black dashed line). The upper-left inset presents the local slope of the MSD, from which the ballistic time scale $T_b$ is determined as the delay time at which the local slope $\xi$ first falls below 95\% of its ballistic regime scaling $\xi=2$ (black dashed line in the inset). The corresponding normalized values $T_b/\tau_\eta$ are shown in the lower-right inset.}
\label{fig.5}
\end{figure}

This expectation is supported by the MSD results in Fig.~\ref{fig.5}. The MSD displays a clear transition from an initial ballistic regime, $\langle |\Delta X|^2 \rangle \sim \tau^2$, to a diffusive regime at longer times, $\langle |\Delta X|^2 \rangle \sim \tau$. Such a crossover is characteristic of particles transported in turbulent flows, where short-time motion is dominated by inertia and long-time dispersion by turbulent mixing \cite{falkovich2001particles,girotto2024lagrangian}. To quantify the transition time, we define it as the earliest instant at which the local slope $\xi = \mathrm{d}\log \langle |\Delta X|^2 \rangle / \mathrm{d}\log \tau$ falls below 95\% of its ballistic regime scaling $\xi=2$, as illustrated in the upper-left inset of Fig.~\ref{fig.5}. The resulting normalized ballistic time scale $T_b/\tau_\eta$, shown in the lower-right inset of Fig. \ref{fig.5}, increases systematically with droplet size. This confirms that larger droplets indeed exhibit an extended ballistic regime, reflecting their enhanced inertia effect in maintaining Lagrangian motion.

\section{Conclusion}
We have experimentally investigated the transport behavior of finite-sized, neutrally buoyant droplets in homogeneous isotropic turbulence, focusing on both their size distribution and Lagrangian dynamics. Using a soccer-ball-type HIT apparatus in conjunction with a droplet-resolved Lagrangian tracking framework, we showed that the dispersed phase exhibits a log-normal size distribution whose mean value and variance decrease with increasing Reynolds number, indicating a progressive trend toward mono-dispersity at higher turbulence intensities. Although the velocity and acceleration PDFs and their corresponding variances display only a modest dependence on the droplet size, the temporal statistics reveal clear size-dependent dynamical effects: larger droplets exhibit slightly longer Lagrangian velocity integral times and a visibly extended ballistic regime in their MSD, consistent with their increasing effective size and inertia. These results suggest that in the droplet size range investigated herein, mildly deformable droplets with internal circulation behave similarly to rigid finite-size particles. However, differences would be expected for larger droplets whose deformation and internal circulation become significant, leading to deviations from the rigid-particle Lagrangian behavior.

Overall, the present findings not only reinforce the established understanding of finite-size particle dynamics but also demonstrate that our experimental facility provides a stable and versatile platform for studying droplet-laden turbulence at low volume fractions. This positions the system as a promising tool for future investigations of droplet deformation, breakup, and droplet-flow interactions in dense emulsions.

\acknowledgments
This work was supported by the NSFC Excellence Research Group Program for ``Multiscale problems in nonlinear mechanics" (no. 12588201), NSFC under Grant No. 12402298, and the New Cornerstone Science Foundation through the New Cornerstone Investigator Program and the XPLORER prize.

\end{document}